\documentclass[preprint,aps,12pt,showpacs,nofootinbib,tightenlines]{revtex4}
\usepackage{amsmath}
\usepackage{amssymb}
\usepackage{epsfig}
\usepackage{graphicx}

\begin{document}
\def \epjc{  Eur. Phys. J. C }
\def \jpg{  J. Phys. G }
\def \mpla{ Mod.Phys.Lett. A }
\def \npb{  Nucl. Phys. B }
\def \plb{  Phys. Lett. B }
\def \prd{  Phys. Rev. D }
\def \prl{  Phys. Rev. Lett.  }
\def \pr{   Phys. Rep. }
\def \rmp{  Rev. Mod. Phys. }

\newcommand{\bequ}{\begin{equation}}
\newcommand{\eequ}{\end{equation}}

\newcommand{\beq}{\begin{eqnarray}}
\newcommand{\eeq}{\end{eqnarray}}

\newcommand{\bsll}{B \to X_{s} \ell^+ \ell^-}
\newcommand{\BRbsll}{Br(B \rightarrow X_s \ell^+ \ell^-)}
\newcommand{\bkll}{B\rightarrow K^{*} \ell^+\ell^-}
\newcommand{\mll}{m_{\ell \ell}^2}
\newcommand{\ssa}{\sin^2\theta_W}
\newcommand{\cca}{\cos^2\theta_W}
\newcommand{\tab}[1]{Table \ref{#1}}
\newcommand{\fig}[1]{Fig.\ref{#1}}
\newcommand{\real}{{\rm Re}\,}
\newcommand{\im}{{\rm Im}\,}
\newcommand{\non}{\nonumber\\ }
\newcommand{\s}{\hat{s}}
\newcommand{\np}{{\rm NP}}
\newcommand{\f}{\frac}
\newcommand{\mh}{ m_{H^+}}

\title{\bf The Neutral Higgs Effects on Rare Decay $\bsll$ in T2HDM}
\author{Lin-xia L\"u $^{a,b}$} \email{lulinxia@email.njnu.edu.cn}
\author{Zhen-jun Xiao $^a$} \email{xiaozhenjun@njnu.edu.cn}
\affiliation{a.  Department of Physics and Institute of
Theoretical Physics, Nanjing Normal University, Nanjing, Jiangsu
210097, P.R.China} \affiliation{b.  Department of Physics, Nanyang
Teacher's College, Nanyang, Henan 473061, P.R.China}
\date{\today}

\begin{abstract}
We calculate the new physics contributions to the branching ratios
of the rare decays $\bsll$ $(\ell=e, \mu)$ induced by neutral
Higgs bosons loop diagrams in the top quark two-Higgs-doublet
model (T2HDM). From the numerical calculations, we find that (a)
the neutral Higgs boson's correction to $\bsll$ decays interferes
constructively with its standard model counterpart, but small in
magnitude;  (b) the neutral Higgs contributions to the branching
ratio of $\bsll$ decay can be neglected safely if their masses are
larger than $100$ GeV and $\tan\beta \leq 40$.
\end{abstract}


\maketitle

\section{Introduction}\label{sec-1}

Flavor changing neutral current (FCNC) induced $B$-meson rare
decays occurred only at the loop level in the Standard Model (SM)
and the fact that their branching ratios are tiny seems to be
confirmed by the present experimental data. Since FCNC processes
strongly depend on virtually exchanged particles, they provide a
test of the SM and strong constraints on the parameter space of
new physics models beyond the SM.

Among various rare B meson decay modes, $B \rightarrow X_s \gamma$
decay has received resounding reception in the interested
theoretical physics community. From the $B \rightarrow X_s \gamma$
decay, only the magnitude of $C_{7\gamma}$ instead of its sign can
be constrained by the relevant data. Recently in
Ref.~\cite{gambino05}, the authors investigated the branching
ratio $\BRbsll$ in the Standard Model or with the reversed sign of
$C_{7\gamma}$, and found that the recent data prefer a SM-like
Wilson coefficient $C_{7\gamma}(m_b)$.

The $B$-meson semileptonic decays $\bsll$ $(\ell=e,\mu)$ are of
special interest because it is amenable to a clean theoretical
description, especially for dilepton invariant masses below the
charm resonances, namely in the range $1{\rm GeV^2} \lesssim \mll
\lesssim 6{\rm GeV^2}$. The calculation of the
next-to-next-to-Leading Order (NNLO) QCD corrections in the SM for
$\bsll$ has been completed
\cite{bobeth99,h0109140,hep0306079,hep0411071,hep0312090}. These
semileptonic decays, on the experimental side, have been measured
by Belle and BaBar \cite{ex0109026,babar-04a,belle-05a}. At the
forthcoming LHC-b or the future super $B$ factory experiments, the
dilepton invariant mass spectrum will be measured precisely, which
will provide strong constraints on the new physics beyond the
Standard Model.

In a previous paper \cite{xiao06}, we studied the new physics
contributions to the $B \rightarrow X_s \gamma$ and $\bsll$ decays
induced by the charge-Higgs loop diagrams, and found that a
charge-Higgs boson with a mass lighter than 200 Gev is clearly
excluded by the data, but a charged Higgs boson with a mass around
or larger than 300 GeV is still allowed. In this paper, we will
concentrate on the calculation of new physics contribution to the
semileptonic decays $\bsll$ $(\ell=e,\mu)$ induced by the loop
diagrams involving the neutral-Higgs bosons appeared in the T2HDM.

This paper is organized as follows. In section II, we briefly
review the top quark two-Higgs-doublet model, then calculate the
new penguin or box diagrams induced by neutral Higgs bosons,
extracting out the new physics parts of the Wilson coefficients in
the T2HDM and giving the related formulae for branching ratio
$\BRbsll$. In section III, we present the numerical results for
the branching ratios of the rare decays $\bsll$ in the SM and the
T2HDM.

\section{Rare decays $\bsll$ in the T2HDM}\label{sec-2}

In this section, we present the basic theoretical framework of the
T2HDM and calculate the new physics contributions to the Wilson
coefficients induced by loop diagrams involving the neutral Higgs
bosons.

The new physics  model considered here is the T2HDM proposed in
Ref.~\cite{das1996} and studied for example in
Refs.~\cite{kiers,kiers62,xiao06,lu06}, which is also a special
case of the 2HDM of type III~\cite{hou1992}. The top quark is
assigned a special status by coupling it to one Higgs doublet that
gets a large VEV, whereas all the other quarks are coupled only to
the other Higgs doublet whose VEV is much smaller. As a result,
$\tan\beta$ is naturally large in this model.

The Yukawa interaction of the T2HDM can be written as
follows~\cite{das1996}:
\beq {\cal L }_Y = - {\overline{L}}_L \phi_1 E l_R -
{\overline{Q}}_L \phi_1 F d_R - {\overline{Q}}_L
{\widetilde{\phi}}_1 G {\bf 1}^{(1)} u_R -
{\overline{Q}}_L{\widetilde{\phi}}_2 G {\bf 1}^{(2)} u_R + H.c.
\eeq
where $\phi_i$ $(i=1,2)$ are the two Higgs doublets with
${\widetilde{\phi}}_i = i \tau_2 \phi^*_i $; and $E$, $F$, $G$ are
the generation space $3 \times 3$ matrices; $Q_L$ and $L_L$ are
3-vector of the left-handed quark and lepton doublets; ${\bf 1}
^{(1)} \equiv diag(1,1,0)$; ${\bf 1} ^{(2)} \equiv diag(0,0,1)$
are the two orthogonal projection operators onto the first two and
the third families respectively.

The Yukawa couplings for quarks are of the form~\cite{das1996}
\begin{align}
{\cal L}_Y  =&-\sum\limits_{D=d,s,b} m_D \bar{D} D -
\sum\limits_{U=u,c,t} m_U \bar{U} U
\nonumber\\
&-\sum\limits_{D=d,s,b}\frac{m_D}{v}\bar{D} D [H^0-\tan\beta h^0]
-i\sum\limits_{D=d,s,b}\frac{m_D}{v}\bar{D}\gamma_5 D
[G^0-\tan\beta A^0]
\nonumber\\
&-\frac{m_u}{v}\bar{u} u[H^0-\tan\beta h^0]- \frac{m_c}{v}\bar{c}
c[H^0-\tan\beta h^0]
\nonumber\\
&-\frac{m_t}{v}\bar{t} t[H^0+\cot\beta h^0]
\nonumber\\
&+i\frac{m_u}{v}\bar{u}\gamma_5 u[G^0-\tan\beta A^0]
+i\frac{m_c}{v}\bar{c}\gamma_5 c[G^0-\tan\beta A^0]
\nonumber\\
&+i\frac{m_t}{v}\bar{t}\gamma_5 t[G^0+\cot\beta A^0]
\nonumber\\
& + \frac{g}{\sqrt{2}M_W} \{ - \overline{U}_L V m_D D_R [G^+ -
\tan \beta H^+ ] +\overline{U}_R m_U V D_L [G^+ - \tan \beta H^+]
\nonumber\\
& + \overline{U}_R \Sigma^ {\dag} V D_L [\tan \beta + \cot \beta]
H^+ +h.c. \}.
\end{align}
where $G^{\pm}$ and $G^0$ are Goldstone bosons, $H^{\pm}$ are
charged Higgs bosons, while the CP-even $(H^0, h^0)$ and CP-odd
$A^0$ are the so-called neutral Higgs bosons. Here $M_U$ and $M_D$
are the diagonal up- and down-type mass matrices, $V$ is the usual
Cabibbo-Kobayashi-Maskawa  ( CKM ) matrix and $\Sigma \equiv M_U
U^{\dag}_R {\bf 1}^{(2)} U_R$. $U^{\dag}_R$ is the unitary matrix
which diagonalizes the right-handed up-type quarks as defined in
Ref.~\cite{kiers}.


The effective hamiltonian inducing the transition $b \rightarrow s
\ell^+ \ell^-$ at the scale $\mu$ has the following
structure~\cite{huang}:
\beq \label{Heff} {\cal H} = - \frac{4 G_F}{\sqrt{2}} V^*_{ts}
V_{tb} \sum \limits_{i=1}^{10} [C_i(\mu) {\cal O}_i(\mu)+
C_{Q_i}(\mu) Q_i(\mu) ] \eeq
Where $C_i$, $C_{Q_i}$ are the Wilson coefficients at the
renormalization point $\mu=m_W$, ${\cal O}_i$'s ($i=1,\cdots,10$)
are the operators in the SM and are the same as those given in the
Ref.~\cite{bobeth99}, and $Q_i$'s come from exchanging the neutral
Higgs bosons in T2HDM and have been given in Ref.~\cite{huang}.
$G_F=1.16639\times 10^{-5}\; GeV^{-2}$ is the Fermi coupling
constant, and $V^*_{ts} V_{tb}$ is the CKM factor. We work in the
approximation where the combination $(V_{us}^* V_{ub})$ of the CKM
matrix elements is neglected. The top-quark and charm-quark
contributions are added up with the results in the summed form.

\begin{figure}[]
\vspace{-7cm}
\centerline{\mbox{\epsfxsize=18cm\epsffile{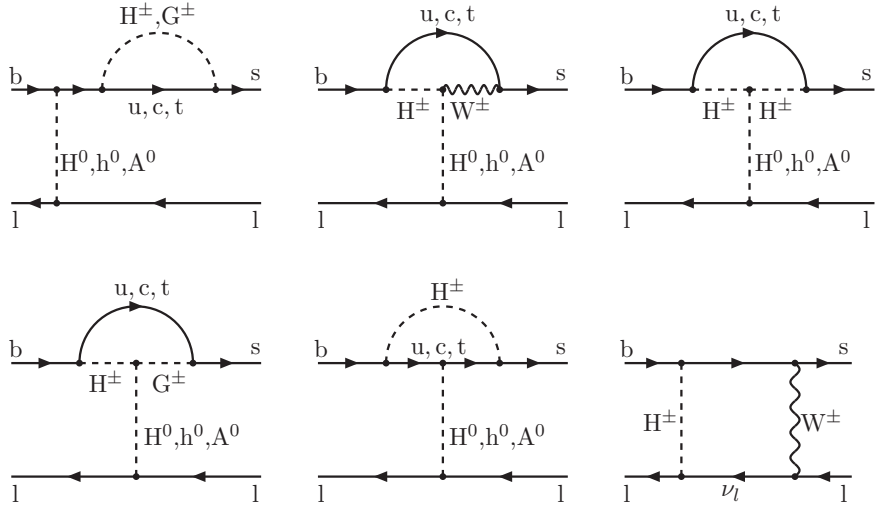}}}
\vspace{-11cm} \caption{ The typical Feynman diagrams for the
decay $\bsll$ when the new physics contributions from the loops
involving the neutral Higgs bosons in T2HDM. The box diagram in
the lower right corner is an example of the diagrams involving the
charged Higgs boson. } \label{fig:fig1} \vspace{20pt}
\end{figure}

In the framework of the SM, the rare decays $\bsll$ proceed
through loop diagrams and are of forth order in the weak coupling.
The dominant contributions to this decay come from the $W$ box and
$Z$ penguin diagrams. The corresponding one-loop diagrams in the
SM were evaluated long time ago and can be found for example in
Refs.\cite{inami1981,misiak1992}.

In the T2HDM, the $\bsll$ decays proceed also via additional loops
involving charged and/or  neutral Higgs bosons exchanges. In
Ref.~\cite{xiao06}, we have given a detailed derivation of the
lengthy expressions of the T2HDM corrections to the relevant
Wilson coefficients induced by the loop diagrams involving charged
Higgs bosons. Here we first consider the neutral Higgs bosons
contributions to the Wilson coefficients.

At the high energy scale $\mu_W\sim M_W$, the leading
contributions to $C_{Q_i}$ come from the diagrams in
Figs.\ref{fig:fig1}. By calculating the Feynman diagrams, we find
analytically that
\begin{eqnarray}
C_{Q_1}(M_W)&=&-f_{ac}\sum \limits_{i=c, t} \kappa^{is} \left\{
\frac{m_i^2}{m_{h^0}^2} \left(-\tan^2\beta + \frac{(\Sigma^T
V^*)_{is}}{m_i V_{is}^*} (\tan^2\beta+1)\right)\bar{B}_0(y_i)
\right. -\frac{m_i^2}{m_{h^0}^2}\bar{B}_0(x_i)
\nonumber \\
& &-\frac{M_W^2}{m_{h^0}^2}\left[ x_i \left(-1+
\frac{(\Sigma^\dagger V)_{ib}}{m_i V_{ib}} (\cot^2\beta+1)
\right)\left(2 \bar{C}_{01}(x_i,y_i,x_{H^+})-
\bar{C}_{11}(x_i,y_i,x_{H^+}) \right) \right.
\nonumber \\
& &+\left. \frac{m_b^2}{M_W^2} \left(2
\bar{C}_{11}(x_i,y_i,x_{H^+})  - \bar{C}_{22}(x_i,y_i,x_{H^+})
\right) + \bar{C}_{21} (x_i,y_i,x_{H^+}) \right]
\nonumber \\
& &+x_i \left(\frac{m_{H^+}^2}{m_{h^0}^2}-1\right) \left[
\left(-1+ \frac{(\Sigma^\dagger V)_{ib}}{m_i V_{ib}}
(\cot^2\beta+1) \right)\bar{C}_{11}(x_i,y_i,x_{H^+})\right. \non
&& \left. +\bar{C}_{01}(x_i,y_i,x_{H^+}) \right]
\nonumber \\
& & + \frac{m_i^2(2m_{H^+}^2+m_{H^0}^2-2m_{h^0}^2)} {m_{H^+}^2
m_{H^0}^2} \left(-1 + \frac{(\Sigma^T V^*)_{is}}{m_i V_{is}^*}
(\cot^2\beta+1)\right)
\nonumber \\
& & \times \left[ \left(-1+ \frac{(\Sigma^\dagger V)_{ib}}{m_i
V_{ib}} (\cot^2\beta+1) \right) C_{11}(y_i)+C_{01}(y_i) \right ]
\nonumber \\
& & +\frac{m_i^2 Q_{h^0}' \tan\beta }{m_{h^0}^2} \left[y_i
\left(-1 + \frac{(\Sigma^T V^*)_{is}}{m_i V_{is}^*}
(\cot^2\beta+1)\right)\right. \non && \left. \times
\left(C_{01}''(y_i) -\frac{m_b^2}{m_i^2}
C_{11}''(y_i)-\frac{1}{y_i}C_{21}''(y_i) \right) \right.
\nonumber \\
& &\left. +y_i \left(-1 + \frac{(\Sigma^T V^*)_{is}}{m_i V_{is}^*}
(\cot^2\beta+1)\right) \left(-1+ \frac{(\Sigma^\dagger
V)_{ib}}{m_i V_{ib}} (\cot^2\beta+1) \right) \right. \non &&
\left. \times  \left( C_{01}''(y_i)-2C_{11}''(y_i)\right) \right.
\non & &\left. \left. +\frac{m_b^2}{18M_{H^+}^2} \left(-1 +
\frac{(\Sigma^T V^*)_{is}}{m_i V_{is}^*} (\cot^2\beta+1)\right)
C_{22}''(y_i)\right] -B_{+}(x_{H^+},x_t) \right\} \; ,
\end{eqnarray}
\begin{eqnarray}
C_{Q_2}(M_W)&=&f_{ac}\sum \limits_{i=c, t} \kappa^{is} \left\{
\frac{m_i^2}{m_{A^0}^2}\left[ \left( -\tan^2\beta +
\frac{(\Sigma^T V^*)_{is}}{m_i V_{is}^*}
(\tan^2\beta+1)\right)\bar{B}_0(y_i)-\bar{B}_0(x_i)\right]
\right. \nonumber \\
& &-\frac{M_W^2}{m_{A^0}^2} \left[ x_i \left(-1+
\frac{(\Sigma^\dagger V)_{ib}}{m_i V_{ib}} (\cot^2\beta+1)
\right)\left(2 \bar{C}_{01}(x_i,y_i,x_{H^+}) -
\bar{C}_{11}(x_i,y_i,x_{H^+})\right) \right.
\nonumber \\
& &+ \left. \frac{m_b^2}{M_W^2}\left(2
\bar{C}_{11}(x_i,y_i,x_{H^+}) -
\bar{C}_{22}(x_i,y_i,x_{H^+})\right) +
\bar{C}_{21}(x_i,y_i,x_{H^+})\right]
\nonumber \\
& &+x_i\left(\frac{m_{H^+}^2}{m_{A^0}^2}-1\right) \left[ \left(-1+
\frac{(\Sigma^\dagger V)_{ib}}{m_i V_{ib}} (\cot^2\beta+1)
\right)\bar{C}_{11}(x_i,y_i,x_{H^+})\right.\non && \left.
+\bar{C}_{01}(x_i,y_i,x_{H^+})\right]
\nonumber \\
& &-\frac{m_i^2 Q_{A^0}' \tan\beta }{m_{A^0}^2} \left[y_i \left(-1
+ \frac{(\Sigma^T V^*)_{is}}{m_i V_{is}^*} (\cot^2\beta+1)\right)
\right.\non && \left. \times \left(C_{01}''(y_i)
+\frac{m_b^2}{m_i^2} C_{11}''(y_i)+\frac{1}{y_i}C_{21}''(y_i)
\right) \right.
\nonumber \\
& &+y_i \left(-1 + \frac{(\Sigma^T V^*)_{is}}{m_i V_{is}^*}
(\cot^2\beta+1)\right) \left(-1+ \frac{(\Sigma^\dagger
V)_{ib}}{m_i V_{ib}} (\cot^2\beta+1) \right)  C_{01}''(y_i)
\nonumber \\
& &-\frac{m_b^2}{18M_{H^+}^2}\left.\left. \left(-1 +
\frac{(\Sigma^T V^*)_{is}}{m_i V_{is}^*} (\cot^2\beta+1)\right)
C_{22}''(y_i)\right] -
B_{+}(x_{H^+},x_t) \right\} \; , \\
C_{Q_3}(M_W)&=&\frac{m_b \; e^2}{m_{\ell} \; g_s^2}(C_{Q_1}(M_W)+C_{Q_2}(M_W))\; , \\
C_{Q_4}(M_W)&=&\frac{m_b \; e^2}{m_{\ell} \; g_s^2}(C_{Q_1}(M_W)-C_{Q_2}(M_W)) \; , \\
C_{Q_i}(M_W)&=&0, ~~~~{\rm for } \ \ i=5,\cdots, 10,
\end{eqnarray}
where $f_{ac}=\frac{m_b m_{\ell} \tan^2\beta}{4 M_W^2 \ssa }$,
$\kappa^{is}=-V_{ib}V_{is}^*/(V_{tb}V_{ts}^*)$,
$x_{H^+}=m_{H^+}^2/M_W^2$, $x_{i}=m_i^2/M_W^2$,
$y_{i}=m_i^2/m_{H^+}^2$, and
$Q_{A^0}'=Q_{h^0}'=\tan\beta(-\cot\beta)$ for $c$ $(t)$ quark. The
one-loop integral functions appeared in $C_{Q_1}(M_W)$ and
$C_{Q_2}(M_W)$ can be written as
\begin{eqnarray}
\bar{B}_0(y)&=&1+\frac{y}{1-y}\ln[y] \; ,
\nonumber \\
B_{+}(x,y)&=&\frac{y}{x-y}\left(\frac{
\ln[x]}{1-x}-\frac{\ln[y]}{1-y}\right) \; ,
\nonumber \\
C_{01}(y)&=&\frac{1}{1-y}+\frac{y}{(1-y)^2}\ln[y]\; ,
\nonumber \\
C_{11}(y)&=&\frac{1-3y}{4(1-y)^2}-\frac{y^2}{2(1-y)^3}\ln[y]\; ,
\end{eqnarray}
\begin{eqnarray}
C_{01}''(y)&=&-\frac{1}{1-y}-\frac{1}{(1-y)^2}\ln[y] \; ,
\nonumber \\
C_{11}''(y)&=&\frac{y-3}{4(1-y)^2}-\frac{1} {2(1-y)^3}\ln[y] \; ,
\nonumber \\
C_{21}''(y)&=&\frac{3-y}{2(1-y)}+\frac{1} {(1-y)^2}\ln[y] \; ,
\nonumber \\
C_{22}''(y)&=&\frac{-11+7y-2y^2}{(1-y)^3}-
\frac{6}{(1-y)^4}\ln[y], \non \bar{C}_{01}(x,y,z)&=&\frac{y
\ln[x]- x \ln[y]
            -\ln[z]} {(1-x)(1-y)(1-z)} \; ,
\nonumber \\
\bar{C}_{11}(x,y,z)&=&-\frac{1}{2(1-y)(1-z)} -
\frac{y^2}{2(1-x)(1-y)^2}
            \ln[y] -\frac{1}{2(1-x)(1-z)^2} \ln[z] \; ,
\nonumber \\
\bar{C}_{21}(x,y,z)&=& \frac{3}{2} - \frac{x y}{(1-x)(1-y)}\ln[y]+
              \frac{1}{(1-x)(1-z)}\ln[z] \; ,
\nonumber \\
\bar{C}_{22}(x,y,z)&=& \frac{-3x+5y+z-3}{6(1-y)^2(1-z)^2} +
              \frac{y^3}{3(1-x)(1-y)^3}\ln[y] -
              \frac{1}{3(1-x)(1-z)^3}\ln[z]\; .
\end{eqnarray}
%


Neglecting the strange quark mass, the effective Hamiltonian
(\ref{Heff}) leads to the following matrix element for the rare
decays $\bsll$
\begin{eqnarray}
{\cal M}&=&\frac{\alpha_{em}G_F}{2\sqrt{2}\pi}V_{tb}V_{ts}^*
\left\{-2\widetilde
C_{7\gamma}^{eff}\frac{m_b}{q^2}\bar{s}i\sigma_{\mu\nu}p_\nu(1+\gamma_5)b\bar{\ell}\gamma_\mu\ell
+ \widetilde C_{9V}^{eff}\bar{s}\gamma_\mu(1-\gamma_5)
b\bar{\ell}\gamma_\mu\ell \right. \non && \left. + \widetilde
C_{10A}^{eff}\bar{s}\gamma_\mu(1-\gamma_5)
b\bar{\ell}\gamma_\mu\gamma_5\ell +
C_{Q_1}\bar{s}(1+\gamma_5)b\bar{\ell}\ell
+C_{Q_2}\bar{s}(1+\gamma_5)b\bar{\ell}\gamma_5\ell \right \}.
\label{matrix}
\end{eqnarray}
with $q$ the momentum transfer.

The Wilson coefficients can be evolved from the electroweak scale
$\mu_W \sim M_W$ down to the low-energy scale $\mu \sim m_b$,
according to the renormalization group equation \cite{hep0411071}.
The mixing of the operators ${\cal O}_i (i=1,2,\cdots,10)$ in the
SM has been studied and the anomalous dimension matrix (ADM) has
been given in
Refs.~\cite{h0109140,hep0306079,hep0411071,hep0312090}. Neglecting
the mixing between ${\cal O}_i(i=1,2,\cdots,10)$ and
$Q_i(i=1,2,\cdots,10)$, the effective Wilson coefficients
including charged Higgs bosons contributions at the low scale
$\mu=m_b$ can be found in Ref.~\cite{xiao06}.

The operators ${\cal O}_i(i=1,\cdots,10)$ and $Q_i(i=3,\cdots,10)$
do not mix into $Q_1$ and $Q_2$ and there is no mixing between
$Q_1$ and $Q_2$ \cite{huang1983}. Therefore, the evolution of the
Wilson coefficients $C_{Q_1}$ and $C_{Q_2}$ is
\beq C_{Q_i}(\mu_b)&=&\eta^{-12/23}C_{Q_i}(M_W) \; , \eeq
where $\eta=\alpha_s(M_W)/\alpha_s(\mu_b)$.

In order to eliminate the large uncertainties due to the factor
$m_b^5$ and the CKM elements appearing in the decay width for
$\bsll$, it has become customary to normalize the decay to the
semileptonic decay rate. The integrated branching ratio in
low-$q^2$ region can be written as~\cite{hep0312090,bobeth05}
\beq Br_{\ell\ell} = Br(\bar{B} \to X_c \ell \nu)
\int_{\hat{s}_a}^{\hat{s}_b} \; R(\hat{s}), \label{eq:bsll} \eeq
where $\hat{s} = q^2/m_b^2$ with $\hat{s}_a =1/m_b^2$ and
$\hat{s}_b =6/m_b^2$, $R(\hat{s})$ is the differential decay rate
for the decay $\bsll$ and has been derived in Ref.~\cite{huang}
\begin{eqnarray}
\label{eq:RR} R(\hat{s})& \equiv &
\frac{\frac{d}{d\hat{s}}\Gamma(b \to s l^+ l^-)}{\Gamma(b
\rightarrow ce\overline{\nu})}  = \frac{\alpha_{em}^2}{4
\pi^2}\left|\frac{V^*_{ts} V_{tb}}{V_{cb}}\right|^2 \frac{
(1-\hat{s} )^2}{f(z)\kappa(z)}
\left(1-\frac{4r}{\hat{s}}\right)^{1/2}D(\hat{s}) ,
\end{eqnarray}
where
\begin{eqnarray}
D(\hat{s})&=&4|\widetilde{C}_7^{eff}|^2(1+\frac{2r}{\hat{s}})
(1+\frac{2}{\hat{s}})+
|\widetilde{C}_9^{eff}|^2(1+\frac{2r}{\hat{s}})(1+2\hat{s})
\nonumber\\
&&+|\widetilde{C}_{10}^{eff}|^2 (1-8r+2\hat{s}+\frac{2r}{\hat{s}})
+12{\rm Re}(\widetilde{C}_7^{eff}\widetilde{C}_9^{eff*})
(1+\frac{2r}{\hat{s}})
\nonumber\\
&&+\frac{3}{2}|C_{Q_1}|^2(\hat{s}-4r)+\frac{3}{2}|C_{Q_2}|^2\hat{s}
+6{\rm Re}(\widetilde{C}_{10}^{eff}C_{Q_2}^*)r^{1/2}.
\label{bsllp}
\end{eqnarray}
Here $r=m_\ell^2/m_b^2$, $z=m_c/m_b$,
$f(z)=1-8z^2+8z^6-z^8-24z^4\ln{z}$ is the phase-factor, and
$\kappa(z) \simeq 1-\frac{2\alpha_s(\mu)}{3\pi} \left[ \left(\pi^2
-\frac{31}{4}\right)(1-z)^2 + \frac{3}{2} \right]$ is the single
gluon QCD correction to the $b \rightarrow ce\bar{\nu}$ decay.

\section{Numerical result}\label{sec-3}

In numerical calculations, we will use the following input
parameters
\beq m_d&=&5.4 MeV,\quad m_s=150 MeV,\quad m_b=4.6 GeV,  \non
m_c&=&1.4 GeV, \quad \overline{m}_t(m_t)=165.9 GeV, \quad
m_{B_d}=5.279 GeV,\quad m_{B_s}=5.367 GeV, \non A&=&0.853, \quad
\lambda=0.225, \quad \bar{\rho}=0.20\pm0.09,\quad
\bar{\eta}=0.33\pm 0.05, \label{eq:input} \eeq
where $A$, $\lambda$, $\bar{\rho}$ and $\bar{\eta}$ are
Wolfenstein parameters of the CKM mixing matrix.

From the data of the radiative decay $B \to X_s \gamma$ and $B^0 -
\bar{B}^0$ mixing, we found strong constraints on the parameter
space of the T2HDM~\cite{xiao06}. Here we will consider these
constraints in our choice for the free parameters of the T2HDM.

On the experimental side, the average of the measured branching
ratios of $\bsll$ $(\ell=e,\mu)$ for the low dilepton invariant
mass region ($1{\rm GeV^2}<m_{\ell\ell}^2\equiv q^2<6{\rm GeV^2}$)
as given in Ref.~\cite{gambino05} is
\beq BR(\bsll) = (1.60 \pm 0.51) \times 10^{-6}.
\label{eq:average} \eeq

At NNLO level, the SM prediction after integrating over the
low-$q^2$ region reads
\beq Br(\bsll) &=& \left ( 1.58 \pm 0.08|_{m_t} \pm 0.07|_{\mu_b}
\pm 0.04|_{CKM} \pm 0.06|_{m_b} + 0.18|_{\mu_W}  \right ) \times
10^{-6} \non &=&\left ( 1.58 \pm 0.13 + 0.18|_{\mu_W}  \right )
\times 10^{-6}. \label{eq:smp} \eeq
where the errors show the uncertainty of input parameters of
$m_t$, $A$, $\bar{\rho}$, $\bar{\eta}$ and $m_b$, and for $m_b/2
\leq \mu_b \leq 2 m_b$. The last error corresponds to the choice
of $\mu_W =120$ GeV, instead of $\mu_W =M_W$. Since here we focus
on the new physics corrections to the branching ratios of $\bsll$
decay, we will take $\mu_W =M_W$ in the following unless stated
otherwise.

\begin{figure}[tb]
\centerline{\mbox{\epsfxsize=12cm\epsffile{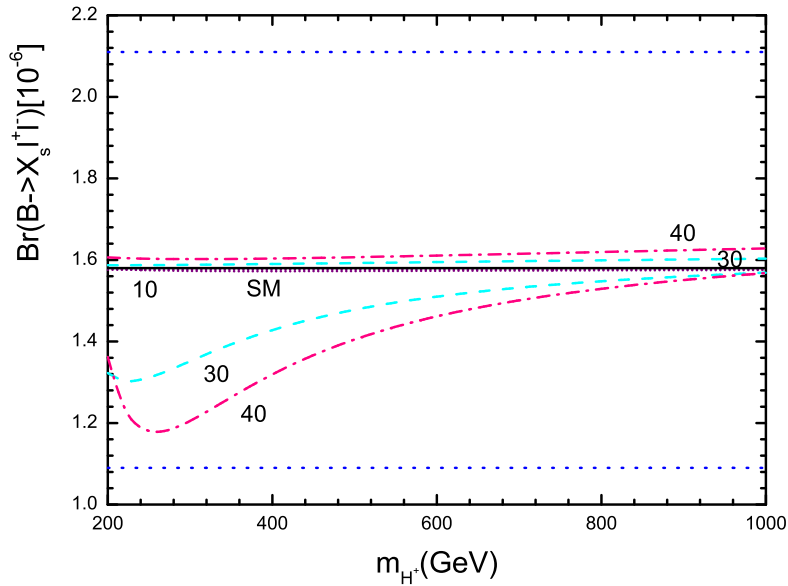} }}
\caption{Plots of the branching ratios of $\bsll$ vs the mass
$m_{H^+}$ in the SM and T2HDM for $\delta=0^\circ$,
$m_{H^0}=160{\rm GeV}$, $m_{h^0}=115{\rm GeV}$, $m_{A^0}=120{\rm
GeV}$ and for $\tan{\beta}=10$, $\tan{\beta}=30$,
$\tan{\beta}=40$, respectively.} \label{fig:fig2}
\end{figure}

\begin{figure}[tb]
\centerline{\mbox{\epsfxsize=12cm\epsffile{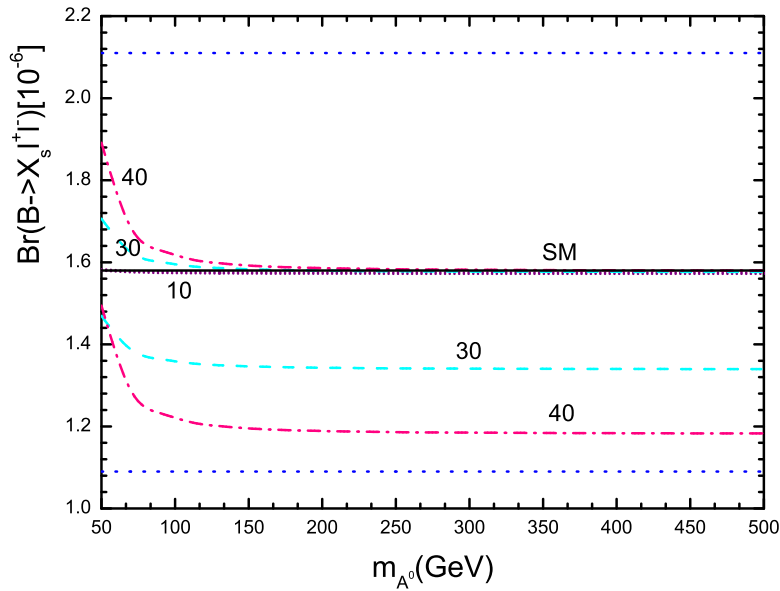} }}
\vspace{0.2cm} \caption{Plots of the branching ratio of $\bsll$ vs
the mass $m_{A^0}$ for $\delta=0^\circ$, $\mh=300{\rm GeV}$,
$m_{H^0}=160{\rm GeV}$, $m_{h^0}=115{\rm GeV}$, and for
$\tan{\beta}=10$, $30$, $40$, respectively. } \label{fig:fig3}
\end{figure}

The new physics corrections to the branching ratio of $\bsll
(\ell=e, \mu)$ in T2HDM are shown in Fig.~\ref{fig:fig2} and
Fig.~\ref{fig:fig3}. The band between two horizontal dot lines
refers to the data within $1\sigma$ error: $Br(\bsll)=(1.60 \pm
0.51) \times 10^{-6}$; while the solid line corresponds to the
central value of the SM prediction at NNLO level:
$Br(\bsll)=1.58\times 10^{-6}$.

In Fig.~\ref{fig:fig2}, the dot-dashed and dashed curve little
above the solid line ( SM prediction) are the T2HDM predictions
for $\tan\beta=40$ and $30$ respectively, when only the new
physics contributions from neutral Higgs bosons are taken into
account (the case A), while the dot-dashed and dashed curves below
the solid line (SM prediction) show the corresponding T2HDM
predictions when the new physics contributions from both the
neutral and charged Higgs bosons are included (the case B). For
$\tan\beta \leq 10$, the new physics contributions in both case A
and B are always very small and can be neglected safely.

In Fig.~\ref{fig:fig3}, we show the the $m_{A^0}$ dependence of
$Br(\bsll)$ for $\delta=0^\circ$, $\mh=300{\rm GeV}$,
$m_{H^0}=160{\rm GeV}$, $m_{h^0}=115{\rm GeV}$, and for
$\tan{\beta}=10$, $30$, $40$, respectively. Again, the dot-dashed
and dashed curve little above ( below) the central solid line are
the T2HDM predictions for the case A ( case B) and for
$\tan\beta=40$ and $30$ respectively. For $\tan\beta \leq 10$, the
curves in the T2HDM can not be separated with the solid line (SM
prediction).

For the CP-even neutral Higgs boson $H^0$ and $h^0$, we have the
similar results. The neutral Higgs bosons contributions to the
decays $\bsll$ are always very small if their masses are heavier
than $100$ GeV as suggested by the direct experimental searches.

To summarize, we have calculated the new physics contributions to
the rare B meson decays $\bsll$ induced by the loop diagrams
involving the neutral or charged Higgs bosons in the top-quark
two-Higgs-doublet model, and compared the theoretical predictions
in the SM and the T2HDM with currently available data. From the
numerical results and the figures, we found the following points
\begin{enumerate}
\item[]{(i)} The neutral Higgs contributions to the branching
ratio $Br(\bsll)$ interfere constructively with their SM
counterparts, but small in magnitude. The charged Higgs, however,
can provide large new physics contribution to both $B \to X_s
\gamma$ and $\bsll$ decays.

\item[]{(ii)} The neutral Higgs contributions to the branching
ratio of $\bsll$ decay can be neglected safely if their masses are
larger than $100$ GeV and $\tan\beta\leq 40$.

\item[]{(iii)} Within the considered parameter space of the T2HDM,
the theoretical predictions for $\BRbsll$ always agree well with
the measured value within one standard deviation.

\end{enumerate}

\begin{acknowledgments}
This work is partly supported by the National Natural Science
Foundation of China under Grant No.10575052, and by the
Specialized Research Fund for the Doctoral Program of Higher
Education (SRFDP) under Grant No.~20050319008.

\end{acknowledgments}

\newpage



\begin{thebibliography}{99}

\bibitem{gambino05}
P.~Gambino, U.~Haisch, and M.~Misiak, \prl {\bf 94} 061803 (2005).

\bibitem{bobeth99}
C.~Bobeth, M.~Misiak and J.~Urban, \npb {\bf 574}, 291 (2000)
[hep-ph/9910220].

\bibitem{h0109140}
H.~H. Asatryan {\it et al}, \prd {\bf 65}, 074004 (2002); H.~H.
Asatryan {\it et al}, \prd {\bf 66}, 034009 (2002); A.~Ghinculov
{\it et al}, \npb {\bf 648}, 254 (2003); H.~M.~Asatrian {\it et
al}, \prd {\bf 66}, 094013 (2002).

\bibitem{hep0306079}
P.~Gambino, M.~Gorbahn and U.~Haisch, \npb {\bf 673}, 238 (2003).

\bibitem{hep0411071}
M.~Gorbahn and U.~Haisch, \npb {\bf 713}, 291 (2005).

\bibitem{hep0312090}
C.~Bobeth, P.~Gambino, M.~Gorbahn and U.~Haisch, JHEP {\bf 0404}
071(2004).

\bibitem{ex0109026}
K.~Abe {\it et al}., [Belle Collaboration], \prl {\bf 88}, 021801
(2002); B.~Aubert {\it et al}. [BaBar Collaboration] \prl {\bf 88}
241801 (2002).

\bibitem{babar-04a}
BaBar Collaboration, B.~Aubert {\it et al}., \prl {\bf 93}, 081802
(2004).

\bibitem{belle-05a}
Belle Collaboration, M.~Iwasaki {\it et al}., \prd {\bf 72} ,
092005 (2005).

\bibitem{xiao06}
Zhen-jun Xiao  and Lin-xia L\"{u}, \prd {\bf 74}  (2006) 034016.

\bibitem{das1996}
A.~Das and C.~Kao, \plb {\bf 372}, 106 (1996).

\bibitem{kiers}
K.~Kiers, A.~Soni and G.H.~Wu, \prd {\bf 59} (1999) 096001;
G.H.~Wu and A.~Soni, \prd {\bf 62} (2000) 056005.

\bibitem{kiers62}
K.~Kiers, A.~Soni and G-H.~Wu, \prd, {\bf 62} (2000) 116004.

\bibitem{lu06}
L.X.~L\"{u} and Z.J.~Xiao, hep-ph/0609279.

\bibitem{hou1992}
W.-S.~Hou, \plb {\bf 296} (1992) 179 ; M.~Luke and M.J.~Savage,
\plb {\bf 307} (1993) 387.

\bibitem{huang}
Yuan-Ben~Dai, Chao-Shang~Huang, and Han-Wen~Huang, \plb {\bf 390}
257(1997); Erratum-ibid, {\bf 513}, 429 (2001); Chao-Shang~Huang,
Liao~Wei, Qi-Shu~Yan, and Shou-Hua~Zhu, \prd {\bf 63} 114021
(2001); Erratum-ibid, {\bf 64}, 059902 (2001).

\bibitem{inami1981}
T.~Inami and C.S.~Lim,  Prog. Theor. Phys, {\bf 65}, 297 (1981)
[erratum {\bf 65}, 1772 (1981)].

\bibitem{misiak1992}
M.~Misiak, \npb {\bf 393}, 23 (1993).

\bibitem{huang1983}
C.S.~Huang, Commun. Theor. Phys. {\bf 2}, 1265 (1983).

\bibitem{bobeth05}
C.~Bobeth, A.J.~Buras, F.~Fr\"uger and J.~Urban, \npb {\bf 630},
87 (2002); C.~Bobeth, A.J.~Buras and T.~Ewerth, \npb {\bf 713},
522 (2005).

\end{thebibliography}
\end{document}